\documentclass{elsart}
\usepackage{epsfig}
\begin{document}
\begin{frontmatter}

\title{All-particle primary energy spectrum in the 3-200 PeV energy range}
\author[ARM]{A.P. Garyaka}, 
\author[ARM]{R.M. Martirosov},
\author[USB]{S.V. Ter-Antonyan\corauthref{cor}},
  \ead{samvel$\_$terantonyan@subr.edu}	
\corauth[cor]{Corresponding author:}
\author[RUS]{A.D. Erlykin},
\author[RUS]{N.M. Nikolskaya},
\author[MON]{Y.A. Gallant},
\author[USA]{L.W. Jones}
\author[FRA]{and J. Procureur}
\address[ARM]{Yerevan Physics Institute, 2 Alikhanyan Br. Str., 
375036 Yerevan, Armenia}
\address[USB]{Department of Physics, Southern University, Baton Rouge, USA}
\address[RUS]{P.N.Lebedev Physical Institute, Moscow, Russia}
\address[MON]{LPTA,
Universit$\acute{e}$ Montpellier II, CNRS/IN2P3, Montpellier, France}
\address[USA]{University of Michigan, Department of Physics, USA}
\address[FRA]{Centre d'Etudes Nucl\'eaires de Bordeaux-Gradignan, 
Gradignan, France}

\parindent=15pt
\begin{abstract}
\indent
We present all-particle primary cosmic-ray energy spectrum
in the $3\cdot10^6-2\cdot10^8$ GeV energy range obtained
by a multi-parametric event-by-event evaluation of the primary energy.
The results are obtained on the
basis of an expanded EAS data set detected at mountain level 
($700$ g/cm$^2$) by the GAMMA experiment.
The energy evaluation method has been developed using
the EAS simulation with the SIBYLL interaction model taking
into account the response of GAMMA detectors and reconstruction
uncertainties of EAS parameters. 
Nearly unbiased ($<5\%$) energy estimations regardless of a
primary nuclear mass with an accuracy of about $15-10\%$ in the 
$3\cdot10^6-2\cdot10^8$ GeV energy range respectively are attained.\\
\indent
An irregularity (`bump') in the spectrum is observed at
primary energies of $\sim7.4\cdot10^7$ GeV.  This bump exceeds a smooth
power-law fit to the data by about 4~standard deviations.
Not rejecting stochastic nature of the bump completely,
we examined the systematic uncertainties of
our methods and conclude that they cannot be responsible for the
observed feature.
\end{abstract}
\end{frontmatter}

\parindent=15pt
\section{Introduction}
\indent
Study of the fine structure in the primary energy spectrum is one 
of the 
most important tasks in the very high energy cosmic ray experiments
 \cite{EW1}. 
Commonly accepted values of the all-particle energy spectrum indexes of 
$-2.7$ and $-3.1$ before and after the knee are an average and may not 
reflect 
the real behavior of the spectrum particularly after the knee. 
It is necessary 
to pay special attention to the energy region of $(1-10)\cdot10^7$ GeV,
where experimental results have been very limited up to now.
Irregularities of the energy spectrum in this region were observed a 
long time ago. They can be seen from energy spectrum obtained more than 
20 years ago with AKENO experiment \cite{AKENO} as well as in later 
works of 
the GAMMA \cite{GPRIM} and TUNKA \cite{TUNKA} experiments.
At the same time the large statistical errors did not allow to discuss 
the reasons of these irregularities. \\
\indent
On the other hand results of 
many experiments on the study of EAS charge particle spectra, 
the behavior of 
the age parameter and muon component characteristics point out that 
the primary mass 
composition at energies above the knee becomes significantly heavier.
Based on these indications, additional investigations of the fine 
structure of 
the primary energy spectrum at $(1-10)\cdot10^7$ GeV have an obvious interest.\\
\indent
There are two ways to obtain the primary energy spectra using
detected extensive air showers (EAS). 
The first way is a statistical method, which unfolds the primary energy spectra
 from the corresponding integral 
equation set based on a detected EAS data set and the model of the EAS
development in the atmosphere \cite{KAS05,KAS07,G5a,G6,G6a}.
The second method is based on an event-by-event evaluation 
\cite{AKENO,G5b,G6b,SC,G05} of the primary energy of 
the detected EAS with parameters
$\mathbf{q}\equiv q(N_e,N_{\mu},N_h,s,\theta)$ using parametric
$E=f(\mathbf{q})$ \cite{AKENO,G5b,G6b,G05} or non-parametric 
\cite{SC} energy estimator previously determined 
on the basis of shower simulations in the framework of a given model 
of EAS development.\\
\indent
Here, applying a new event-by-event parametric energy evaluation 
$E=f(\mathbf{q})$, the all-particle energy spectrum 
in the knee region is obtained on the basis of the 
data set obtained with the GAMMA EAS array \cite{G5a,G6,G6a,G6b} and
a simulated 
EAS database obtained using the SIBYLL \cite{SIBYLL} interaction model.  
Preliminary results have been presented in \cite{G5b,G6b}.    
\section{GAMMA experiment} 
\indent
The GAMMA installation \cite{G5a,G6,G6a,G6b}
is a ground based array of 33 surface 
detection stations and 150 underground muon detectors,
located on the south side of Mount Aragats in Armenia.
The elevation of the GAMMA facility is 3200 m above sea level,
which corresponds to 700 g/cm$^2$ of atmospheric depth. A
diagrammatic layout of the array is shown in Fig.~1.\\
\begin{figure} 
\vspace{20mm}
\begin{center}
\includegraphics[width=12cm, height=6cm]{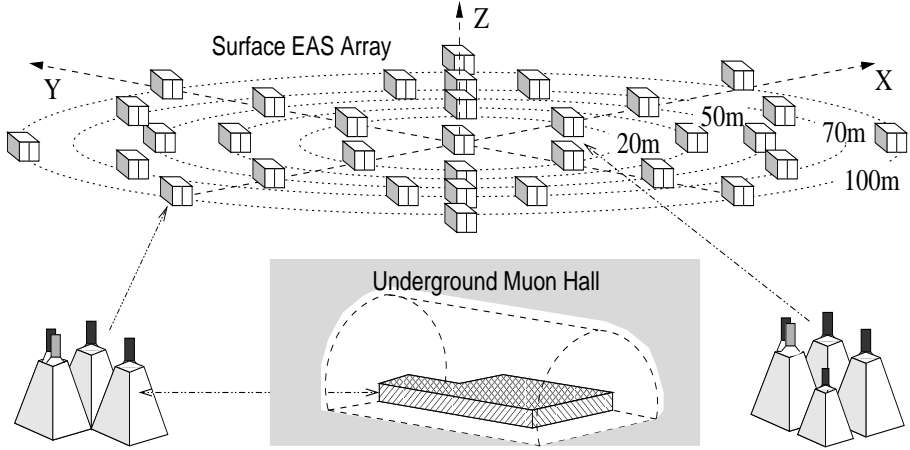}
\end{center}
\caption{Diagrammatic layout of the GAMMA facility.}
\vspace{10mm}
\end{figure}
\indent
The surface stations of the EAS array are arranged in 5 concentric
circles of $\sim$20, 28, 50, 70 and 100 m radii, and each station 
contains
3 plastic scintillation detectors with the dimensions 
of $1\times1\times0.05$ m$^3$. 
Each of the central 9 stations contains an additional (the 4th) 
small scintillator 
with dimensions of $0.3\times0.3\times0.05$ m$^3$ (Fig.~1) for high
particle density ($\gg10^2$ particles/m$^2$) measurements.\\
\indent
A photomultiplier tube is placed on the top of the aluminum
casing covering each scintillator. One of three detectors of each 
station is viewed by two photomultipliers, one of which is designed 
for fast timing measurements.\\
\indent
150 underground muon detectors ('muon carpet') are compactly arranged 
in the 
underground hall under 2.3 kg/cm$^2$ of concrete and rock. 
The scintillator 
dimensions, casings and photomultipliers are the same as in the EAS
surface detectors.\\
\indent
The shower size thresholds
of the $100\%$ shower detection efficiency are equal to
$N_{ch}=3\cdot10^5$ and $N_{ch}=5\cdot10^5$ at the EAS core location
within $R<25$ m and $R<50$ m respectively \cite{G5a,G6,G6b}.\\
\indent
The time delay is estimated by the pair-delay method 
\cite{MAK} to give the time resolution of about $4-5$ ns.
The EAS detection efficiency ($P_d$) and corresponding 
shower parameter reconstruction errors are equal to: 
$P_d=100\%$, $\Delta\theta\simeq1.5^0$, 
$\Delta N_{ch}/N_{ch}\simeq0.05-0.15$, 
$\Delta s\simeq0.05$, 
$\Delta x$ and $\Delta y\simeq0.7-1$ m. 
The reconstruction errors of the truncated muon shower sizes
for $R_{\mu}<50$ m from the shower core
are equal to $\Delta N_{\mu}/N_{\mu}\simeq0.2-0.35$
at $N_{\mu}\simeq10^5-10^3$ respectively \cite{G6,G6a,G6b}.\\
\indent
\section{Event-by-event energy estimation} 
\subsection{Key assumptions}
\indent
Suppose that $E_1=f(\mathbf{q})$ is an estimator of 
energy $E_0$ of unknown primary nuclei which induced showers with 
the detected parameter $\mathbf{q}\equiv q(N_{ch},N_{\mu},s,\theta)$. 
Then the expected all-particle energy spectrum $F(E_1)$ is defined by
\begin{equation}
F(E_1)=
\int\Im(E_0)W(E_0,E_1)\d E_0\;,
\end{equation}
where $\Im(E_0)$ are the energy spectrum of primary nuclei
and $W(E_0,E_1)$ are the corresponding
($E_0,E_1$) transformation probability density function.\\
\indent
If $\Im(E_0)\propto E_0^{-\gamma}$ and $W(E_0,E_1)$ are the 
log-normal distributions with  
$\delta=\overline{E_1/E_0}$ and $\sigma$ 
parameters, the expression (1) has the analytic solution 
for the expected 
spectrum of the energy estimator \cite{Murz}:\\
\begin{equation}
F(E_1)=\Im(E_0)\delta^{\gamma-1}
\exp\Big(
\frac{((\gamma-1)\sigma)^2}{2}
\Big).
\end{equation}  
\indent
It is seen that evaluation of energy spectrum $\Im(E_0)$ from (2)
is possible to perform only at {\em{a priori}} known 
$\gamma$, $\delta$ and $\sigma$ parameters and spectral slope 
($\gamma$)
of detected energy spectra $F(E_1)$ coincides with spectral slope
of primary energy spectra $\Im(E_0)$. 
The values of 
$\delta$ and $\sigma$ may depend on the primary energy
($E_0$) and mass of primary nuclei ($A$) from which the all-particle 
energy spectrum  
$\Im(E_0)=\sum_A\Im_A(E_0)$ is consisted of. 
In this case, the expression (1) is unfolded numerically 
and the slope of detected energy spectrum can differ from primary 
energy spectrum.\\
\indent
For example, the dependence $\sigma(E_0)=a\ln{(E_0/E_{0,\min})}+b$ 
at $|a|<0.1$ leads to the numerical solutions which can be
 approximated
by the expression (2) replacing 
$\sigma$ with $\sigma(E_0)-a\sqrt{\gamma}$. The corresponding
approximation errors
is about $2-5\%$ in the energy range of $E_{\min}-500E_{\min}$ and 
$\gamma\simeq2.3-3.2$.\\
\indent
However, the evaluation of energy spectra can be simplified
provided 
\begin{equation}
\gamma(E_0)\simeq\gamma\pm\Delta\gamma,
\end{equation}
\begin{equation}
\delta(E_0)\simeq\delta_A(E_1)\simeq\delta\equiv1\pm\Delta\delta(E_1),
\end{equation}
\begin{equation}
\sigma(E_0)\simeq\sigma\pm\Delta\sigma
\end{equation}
are satisfied for given energy range of $E_1$. Then, the all-particle
energy spectrum can be evaluated from
\begin{equation}
\Im(E_0)=F(E_1)
\exp\Big(-
\frac{((\gamma-1)\sigma)^2}{2}
\Big)\;.
\end{equation}
\indent
The corresponding error of evaluation (6) with approaches (3-5) 
is determined by a sum of the statistical errors $\Delta F(E_1)$
and systematic errors $\eta$ due to approaches being used: 
\begin{displaymath}
\Big(\frac{\Delta\Im}{\Im}\Big)^2\simeq
\Big(\frac{\Delta F}{F}\Big)^2+\eta^2\;,
\end{displaymath}
where the systematic relative errors $\eta$ is 
\begin{equation}
\eta^2\simeq(\Delta\delta(\gamma-1))^2+
[(\sigma(\gamma-1))^2
(\frac{\Delta\sigma}{\sigma}+
 \frac{\Delta\gamma}{\gamma-1})]^2.
\end{equation}
The values of $\gamma$, $\delta$ and $\sigma$ parameters 
from expressions (3-7) and corresponding uncertainties 
$\Delta\gamma$, $\Delta\delta$ and $\Delta\sigma$ 
essential for the reconstruction of primary energy spectrum
using the GAMMA facility EAS data and approach (6) are
considered in Sections 3.2-3.4 below.
  
\subsection{Uncertainty of spectral slope}
\indent
The results of different experiments 
\cite{G6a,EAS-TOP,CASA,TIBET1} and theoretical
predictions \cite{PB,Hill,Peters} indicate
that the elemental energy spectra 
can be presented in the power law form 
\begin{equation}
\Im_A(E)\propto(\frac{E}{E_k})^{-\gamma},
\end{equation}
where $\gamma=\gamma_1\simeq2.7^{+0.05}_{-0.1}$ at $E<E_k(A)$ 
and $\gamma=\gamma_2\simeq3.15^{+0.1}_{-0.05}$ at $E>E_k(A)$.
It is also accepted that the mass spectra of primary nuclei
can be divided into separate nuclear groups and below,
as in \cite{G6a},
just 4 nuclear species ($H$, $He$, $O$-like and $Fe$-like)
are considered.      
Dependence of the knee energy $E_k$ on the primary nuclei type
assumed to be either rigidity dependent, $E_k=ZE_{Z=1}$
\cite{G6a,PB,Hill,Peters}
or $A$-dependent \cite{G6a,HO}, $E_k=AE_{A=1}$, where $Z$ and $A$
are the charge and  mass of primary nuclei correspondingly.\\
\indent
As a result, the all-particle energy spectrum $\sum_A\Im_A(E)$
slowly changes its slope and can be roughly approximated
by a power law spectrum with power index $\gamma\simeq2.7$
at $E<3\cdot10^6$ GeV,
$\gamma\simeq2.9$ at $3\cdot10^6<E<10^7$ GeV  and
$\gamma\simeq3.1$ at $E>10^7$ GeV. 
This presentation of the all-particle spectrum agrees 
with world data \cite{HO}
in the $\Delta\gamma\simeq0.1$ range of uncertainty and
energy interval  $10^6<E<2\cdot10^8$ GeV.\\
\indent
The values of $\Delta\delta_A(E)$ and $\sigma_A(E_0)$ parameters
are presented in Section~3.4 and depend on efficiency 
of energy estimator $E_1=f(N_{ch},N_{\mu},s,\theta)$.\\
\indent
Notice that it follows from the expression (7)
that for $\sigma\simeq0.1-0.15$ 
and $\Delta\sigma=0.03$
the contribution of $\Delta\gamma$ in the systematic errors (7)
is negligible and the difference of 
all-particle spectra evaluated by expression (6) 
for $\gamma=2.7$ and $\gamma=3.1$ is less than $2\%$ 
at $\sigma\simeq0.15$.\\  
\subsection{The simulated EAS database}
\indent
To obtain the parametric representation for unbiased
($\delta\simeq1$) energy estimator $E_1$ 
of the primary energy $E_0$ we simulated showers database
using the CORSIKA(NKG) EAS simulation code \cite{CORSIKA} 
with the SIBYLL \cite{SIBYLL} interaction model
for $H$, $He$, $O$ and $Fe$ primary nuclei.\\
\indent
Preliminary, the showers simulated with 
NKG mode of CORSIKA code for each of the primary nuclei were compared 
with the corresponding simulations using EGS mode of CORSIKA 
\cite{CORSIKA}
taking into account the detector response, contribution of EAS 
$\gamma$-quanta and shower parameter reconstruction uncertainties.
Simulated statistics were equal to 200 events for each of primary
 nuclei with log-uniform primary energy distribution 
in the range of $2\cdot10^6-10^8$ GeV.
Using the threshold energy of shower electrons (positrons) 
for NKG mode 
at observation level as a free parameter (the same as it was
performed in \cite{G6a}), the biases  
$\delta(N_{ch},A)=(N_{ch}(NKG)/N_{ch}(EGS))-1)$ 
and $\delta_s(A)=s(NKG)-s(EGS)$ were minimized for all simulated 
primary nuclei 
($H,He,O,Fe$).\\
\indent
Applied method of calibration of the NKG mode of CORSIKA
for the GAMMA EAS array differed from 
\cite{G6a} only by the expanded range of selected shower core 
coordinates ($R<50m$) and zenith angles $\theta<45^0$. The obtained
 biases of shower size $\delta(N_{ch})$ and age parameter $s$
in the range of statistical errors ($<5\%$) agreed with data 
\cite{G6a}. The values of $\delta(N_{ch})$ were used further for
correction of the shower size obtained by NKG simulation mode.\\
\indent
The simulated primary energies ($E_0$) for
 shower database were distributed according to a power law 
spectrum $I(E_0)\propto E_0^{-1.5}$ with  
$\mathcal{N}=2\cdot10^4$ total number
of detected ($N_{ch}>5\cdot10^5$, $R<50$m) and reconstructed
showers for each primary nucleus. 
The energy thresholds of primary nuclei
were set as 
$E_{0,\min}(A)\equiv10^6$ GeV and $E_{\max}=5\cdot10^8$ GeV.
The simulated 
showers had core coordinates distributed uniformly within the radius 
of $R<75$ m and zenith angles $\theta<45^0$.\\
\indent
The reconstruction errors $\sigma(\ln{N_{ch}})$ of shower size $N_{ch}$
are presented in Fig.~2 for different primary nuclei and different
zenith angles. 
The right and left ends of diagonals of the rectangular
in Fig.~2 show the average primary energies (in units of GeV) 
responsible for corresponding shower
sizes for the primary proton and Iron nuclei respectively.\\
\begin{figure} 
\begin{center}
\includegraphics[width=7cm]{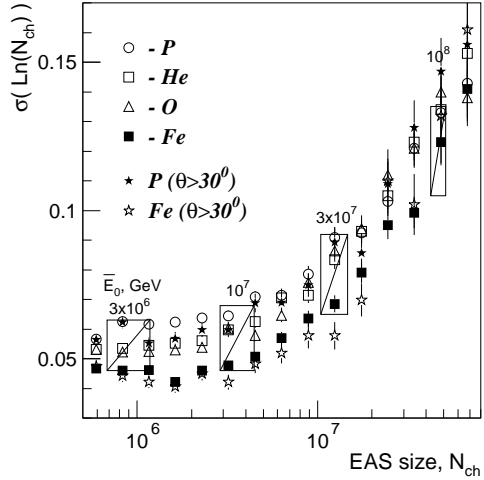}
\end{center}
\caption{Shower size reconstruction errors for different primary 
nuclei ($p,He,O,Fe$) and zenith angles 
($\theta<45^0$ and $30^0<\theta<45^0$). The right and left 
ends of diagonals of the rectangular
show the average primary energies ($\overline{E_0}$) and
corresponding shower sizes computed
for the primary proton and Iron nuclei respectively.}
\vspace{10mm}
\end{figure}
\indent
All EAS muons with energies of $E_{\mu}>4$ GeV at GAMMA observation
level have passed through the  
2.3 kg/cm$^2$ of rock to the muon scintillation carpet (the underground 
muon hall, Fig.~1).
The muon ionization losses and electron (positron)
accompaniment due to muon electromagnetic and photonuclear  
interactions in the rock are taken into account using the  
approximation 
for equilibrium accompanying charged particles obtained from preliminary
simulations with the FLUKA code \cite{FLUKA} in the $0.005-20$ TeV muon
energy range. 
The resulting
charged particle accompaniment per EAS muon in the underground hall
is equal to $0.06\pm0.01$ ($100\%e$) and $11.0\pm1.5$
($98.5\%e,1.4\%h,0.04\%\mu$) at muon energies $0.01$
TeV and $10$ TeV respectively.\\
\indent
Due to absence of saturation in the muon scintillation carpet,
the reconstruction 
errors ($\Delta\ln{N_{\mu}}$) of truncated muon size $N_{\mu}$ 
are continuously 
decreasing with increasing muon truncated sizes in the range
$10^3<N_{\mu}<10^5$. 
Corresponding magnitudes of reconstruction errors
for primary protons and Iron nuclei were equal to 
$\Delta(\ln{N_{\mu,p}})\simeq0.35,0.18,0.15$
and $\Delta(\ln{N_{\mu,Fe}})\simeq0.38, 0.19, 0.10$ 
for EAS muon truncated
sizes $N_{\mu}\simeq10^3, 10^4, 10^5$ respectively.\\
\indent
Fluctuations of the shower size for given primary energies 
$E_{0,A}\equiv10^6,10^7,10^8$ GeV and $\cos{\theta}<0.95$
were equal to $\sigma_{A\equiv p}(N_{ch},E_0)\simeq0.20, 0.14, 0.10$ and
$\sigma_{A\equiv Fe}(N_{ch},E_0)\simeq0.16, 0.13, 0.08$ respectively.\\
\indent
Corresponding fluctuations of muon truncated size were equal to
$\sigma_{A\equiv p}(N_{\mu},E_0)$ $\simeq0.25,0.23,0.2$ and 
$\sigma_{A\equiv Fe}(N_{\mu},E_0)\simeq0.13,0.10,0.08$. For zenith
angles of primary nuclei $45^0>\theta>30^0$, the fluctuations are increased 
about $1.5-2$ times due to the aging of detected showers.\\
\indent
The $4\times2\cdot10^4$ EAS simulated events
with reconstructed $N_{ch}$, $N_{\mu}(R<50$m$)$, 
$s$ and $\theta$ shower parameters for the $E_0$ and $A$
parameters of primary nuclei made up the simulated EAS database. 
\subsection{Energy estimator}
\indent
The event-by-event reconstruction of primary all-particle energy
spectrum using the GAMMA facility is mainly based on high
correlation of primary energy $E_0$ and shower size ($N_{ch}$). 
The shower age parameter ($s$)
zenith angle ($\theta$) and muon truncated shower size 
($N_{\mu}$) have to decrease
the unavoidable biases of energy evaluations due to abundance of
different primary nuclei. In Table~1 the
correlation coefficients $\rho(\mathbf{q},\ln{E_0})$ and 
$\rho(\mathbf{q},\ln{A})$ between shower parameters
$N_{ch}$, $N_{\mu}$, $s$ and primary energy ($E_0$)
and mass of primary nuclei ($A\equiv1,4,16,56$) are presented.\\
\begin{table}  
\caption{
Correlation coefficients $\rho(\mathbf{q},\ln{E_0})$ and
 $\rho(\mathbf{q},\ln{A})$ between shower parameter 
$\mathbf{q}\equiv q(N_{ch},N_{\mu},s)$ and
primary energy ($\ln{E_0}$) and nuclei mass 
$\ln{A}$ for two zenith angular intervals.}
\begin{center}
\begin{tabular}{|c|c|c|c|c|}
\hline
$\mathbf{q} 
$&$\ln{E_0},(\theta<10^0)$&$\ln{E_0},(\theta<45^0)$&$\ln{A},
(\theta<10^0)$&$\ln{A},(\theta<45^0)$\\
\hline
\hline
$\ln{N_{ch}}$&$0.986\pm0.001$&$0.954\pm0.0004$&$0.013
\pm0.013$&$0.007\pm0.004$\\
$\ln{N_{\mu}}$&$0.978\pm0.001$&$0.969\pm0.0003$&$0.139\pm0.012$
&$0.132\pm0.004$\\
$s$&$-0.029\pm0.013$&$-0.02\pm0.004$&$0.018\pm0.013$&$0.015\pm0.004$\\
\hline
\end{tabular}
\end{center}
\vspace{10mm}
\end{table}
\indent
Parametric representation for the energy estimator 
$\ln{E_1}\simeq f(\mathbf{a}|N_{ch},N_{\mu},s,\theta)$ 
we obtained by minimizing $\chi^2$ 
\begin{equation}
\chi^2=\sum_A\sum_{i=1}^{\mathcal{N}}
\frac{(\ln{E_{0,A,i}}-\ln{E_{1,i}})^2}
{\sigma^2}
\end{equation}
with respect to $\mathbf{a}\equiv a(a_1,a_2,\dots,a_{p})$
for different empirical 
functions $f(\mathbf{a})$ with
a different number ($p$) of unknown parameters.
The values of 
$A$, $E_0$ and corresponding reconstructed shower parameters
$N_{ch}$, $N_{\mu}$, $s$ and $\theta$ for estimation of
$E_1$ were taken from simulated EAS database (Section 3.3).\\ 
\indent
The best energy estimations as a result of the 
minimization (9) were achieved 
for the 7-parametric ($p=7$) fit:
\begin{equation}
\ln{E_1}=a_1x+\frac{a_2\sqrt{s}}{c}+a_3+a_4c
+ \frac{a_5}{(x-a_6y)}+a_7ye^s,
\end{equation}
where $x=\ln{N_{ch}}$, $y=\ln{N_{\mu}(R<50m)}$,
$c=\cos{\theta}$, $s$ is the shower age and energy $E_1$ is in GeV.
The values of $a_1,\dots,a_7$ parameters are shown in 
Table~2 and were derived at $\sigma=0.14$  and
$\chi^2_{\min}/n_{d.f.}\simeq1$, where 
the number of degrees of freedom 
$n_{d.f.}=8\cdot10^4$. 
The expected errors $\Delta a_1,\dots,\Delta a_7$ of 
corresponding parameters were negligibly small ($\leq5\%$)
due to very high values of $n_{d.f.}$.\\
\begin{table}  
\caption{
Approximation parameters $a_1,\dots,a_7$ of primary energy evaluation
(10) obtained from $\chi^2$-minimization (9) for the SIBYLL
interaction model, $\sigma=0.14$ and $\chi^2_{\min}/n_{d.f.}\simeq1$.}
\begin{center}
\begin{tabular}{|c|c|c|c|c|c|c|}
\hline
$a_1$&$a_2$&$a_3$&$a_4$&$a_5$&$a_6$&$a_7$\\
\hline
$1.030$&$3.641$&$-5.743$&$2.113$&$6.444$&$1.200$&$-0.045$\\
\hline
\end{tabular}
\end{center}
\end{table}
\indent
The corresponding average biases versus energies 
($E\equiv E_0$ and $E\equiv E_1$)
for the primary proton ($p$), iron ($Fe$) nucleus and uniformly mixed 
$p,He,O,Fe$ composition are presented in Fig.~3 (symbols). 
The boundary 
lines correspond to approximations 
$\Delta\delta\simeq b/\sqrt{E/10^6GeV}$, 
where
$b\simeq0.10$ and $b\simeq-0.17$ for the upper and lower limits 
respectively.
The shaded area corresponds to $b\simeq0.09$ and $b\simeq-0.15$ and were
used to estimate the errors according to (7) for the reconstruction of 
all-particle energy spectrum (Section~4).\\
\indent
The dependence of standard deviations $\sigma(E_0)$  
of systematic errors of energy evaluations (10)  
on primary energy $E_0$ is presented in Fig.~4 for 4 primary nuclei
and uniformly mixed composition with equal fractions
of {\it p}, {\it He}, {\it O} and {\it Fe} nuclei.
\begin{figure} 
\begin{center}
\includegraphics[width=7cm]{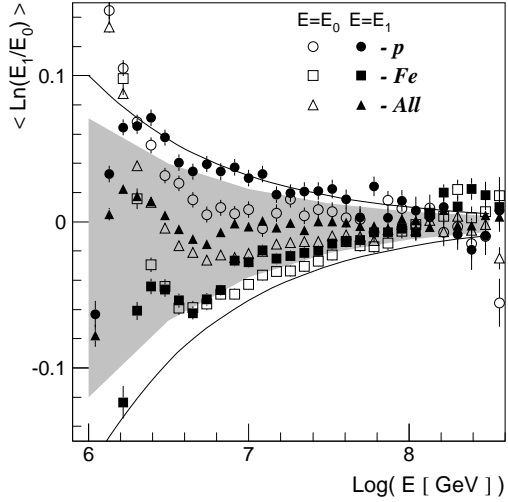}
\end{center}
\caption{Mean biases versus energy $E\equiv E_0$ and $E\equiv E_1$
for the primary proton ($p$) and iron ($Fe$) nucleus and the uniformly
 mixed
$p,He,O,Fe$ compositions ({\em{All}}).}
\label{bias}
\vspace{10mm}
\end{figure}
\begin{figure} 
\begin{center}
\includegraphics[width=7cm]{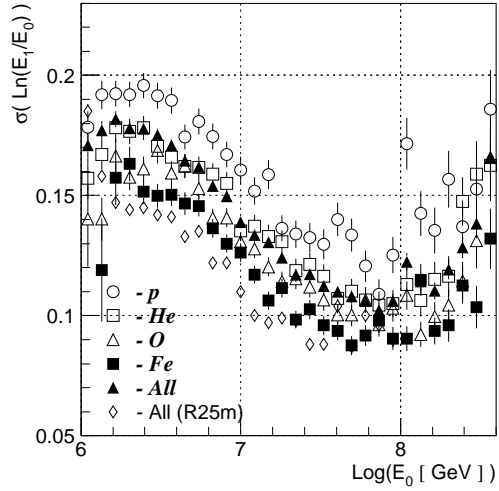}
\end{center}
\caption{Errors of the energy estimator (6)
versus primary energy $E_0$ for 4 primary nuclei
and uniformly mixed ({\em{All}}) composition.
The empty rhombic symbols are taken from our previous data \cite{G5b}
computed for the mixed composition and shower core selection 
criteria $R<25$m.}
\label{sigma}
\vspace{10mm}
\end{figure}
\begin{figure} 
\begin{center}
\includegraphics[width=8cm]{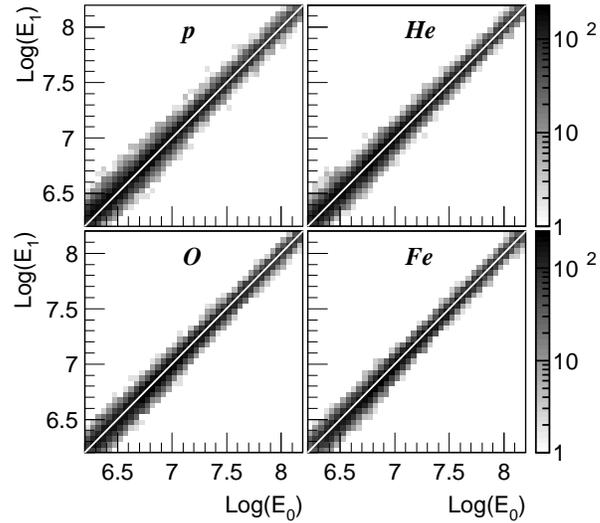}
\end{center}
\caption{$E_0-E_1$ (in units of GeV) scatter plots for four
($p,He,O,Fe$) primary nuclei. 
The white lines show the corresponding $E_0=E_1$ dependence.}
\vspace{10mm}
\end{figure}
\begin{figure} 
\begin{center}
\includegraphics[width=7cm]{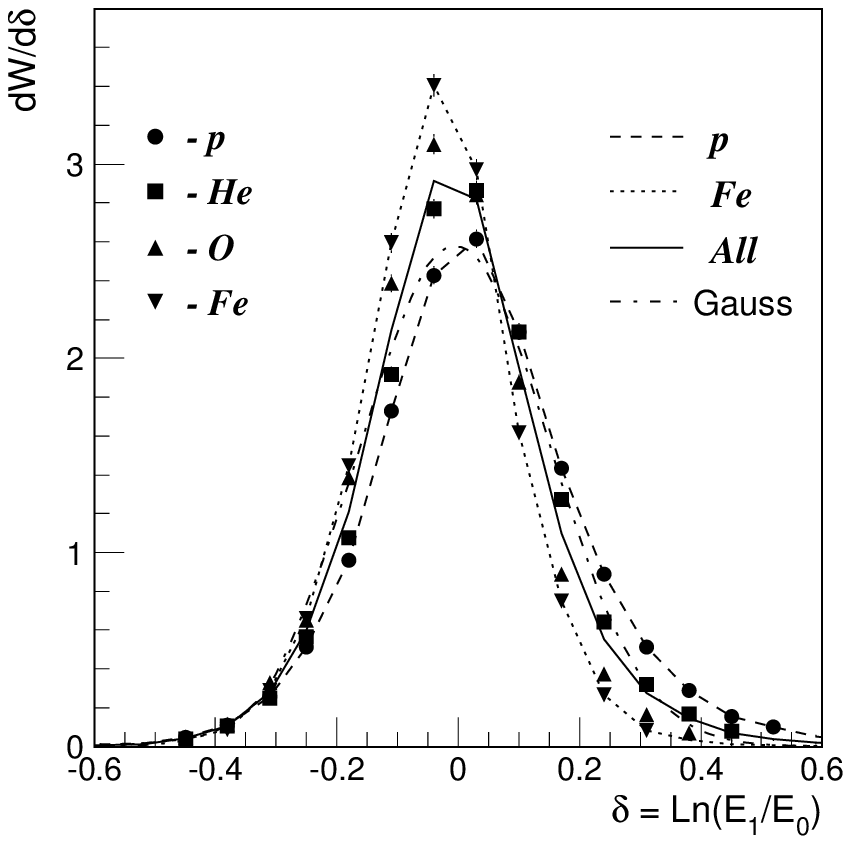}
\end{center}
\caption{Distribution functions of errors for different 
primary nuclei (symbols)
and uniformly mixed composition (symbols and solid line).}
\label{distr}
\vspace{10mm}
\end{figure}
The results for the uniformly mixed composition
with the shower core selection of $R<25$ m \cite{G5b} are presented in 
Fig.~4, as well.
It is seen that the value of $\sigma=0.14$ responsible
for $\chi^2\simeq1$ (expression (9)) 
with uncertainty $\Delta\sigma\simeq0.03$ (expression (5))
encloses the $\sigma_A(E_0)$ data presented in Fig.~4.\\
\indent
Such high accuracies of the energy evaluation
regardless of primary nuclei is a consequence of the high
mountain location of the GAMMA facility (700 g/cm$^2$), 
where the correlation of primary energy with detected EAS size
is about $0.95-0.99$ (Table~1).\\
\indent
The $E_0-E_1$ scatter plot of simulated
primary energy $E_0$ and estimated energy 
$E_1(N_{ch},N_{\mu},s,\theta)$
according to expression (10) and Table~2 are shown in Fig.~5.
The corresponding distributions of energy errors or the kernel
function $W_A(E_0,E_1|\delta_A,\sigma_A)$ of integral equation (1) 
for different primary nuclei and 
uniformly mixed composition are presented in Fig.~6 (symbols).
The average values
$\delta_A(E_0)$ and standard deviations $\sigma_A(E_0)$ 
of these distributions depending on energy
of primary nucleus ($A$) are presented in Figs.~3,4.
The dashed line is an example
of log-normal distribution  
with $\delta$ and $\sigma$ parameters corresponding 
to the uniformly mixed composition.\\
\indent 
It is seen, that the errors can be described by the log-normal 
distributions and
the key assumptions (3-5) are approximately valid.

\indent
The test of applied approaches (expressions (3-6), Section 3.2)
for the reconstruction of all-particle primary energy spectrum was
carried out by the direct folding of
the power law energy spectrum 
$\Im(E_0)\equiv dF_0/dE_0$ (expression (8)) with the log-normal 
kernel function $W(E_0,E_1|\delta(E_0),\sigma(E_0))$ 
according to expression (1) for primary proton and Iron nucleus.
The values of $\delta(E_0)$ and $\sigma(E_0)$ were derived from
the log-parabolic interpolations of corresponding dependencies
presented in Figs.~3,4. 
The event-by-event reconstructed energy spectrum $dF_1/dE_1$ 
(the left hand side of expression (1))
was obtained from expression (6) using approaches (3-5)
with $\sigma=0.14$ and $\Delta\sigma=0.03$. The boundary lines
of shaded area in Fig.~3 were used as estimations  of 
uncertainties $\Delta\delta(E_1)$ of condition (4).
\indent
In Fig.~7 the values of $(dF_1/dE_1)/(dF_0/dE_0)$ 
are presented (symbols) for primary proton and Iron nuclei and
different "unknown" spectral indices of primary energy spectra
(8)
with the rigidity dependent knee at $E_k=3\cdot10^6$ GeV. The shades
area is the expected errors computed according to expression (7).
\begin{figure} 
\begin{center}
\includegraphics[width=8cm]{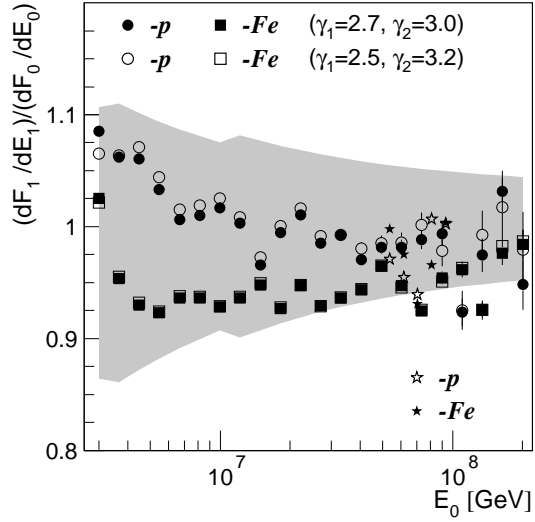}
\end{center}
\caption{
Discrepancies of initial ($dF_0/dE_0$) and reconstructed 
($dF_1/dE_1$) 
energy spectra (symbols) for the different primary
nuclei and spectral indices of initial spectrum. 
The shaded area shows the expected
errors according to expression (7).
The star symbols are the spectral discrepancies for a pulsar
component (Section 5).}
\vspace{10mm}
\end{figure}
\indent
It is seen, that all spectral discrepancies are practically covered
by the expected errors according to expression (7). 
The star 
symbols in Fig.~7 represent the discrepancies of singular spectra 
with knee at energy $7.4\cdot10^7$ GeV described in Section~5.
\section{All-particle primary energy spectrum}  
\indent
The EAS data set analyzed in this paper has been obtained
for $5.63\cdot10^7$ sec of live run time of the GAMMA
facility, from 2004 to 2006.
Showers to be analyzed were selected with the following
criteria: $N_{ch}>5\cdot10^5$, $R<50$ m, 
$\theta<45^\circ$, $0.3<s<1.6$, $\chi^2(N_{ch})/m<3$ and  
$\chi^2(N_{\mu})/m<3$ (where $m$ is the number of scintillators
with non-zero signal),
yielding a total data set of $\sim7\cdot10^5$ selected showers.
The selected measurement range  
provided the $100\%$ EAS detection efficiency 
and similar conditions for the reconstruction of showers 
produced by primary nuclei $H,He,\dots,Fe$ with energies 
$3\cdot10^6<E<(2-3)\cdot10^8$ GeV.  
The upper energy limit is determined from
Fig.~4, where the saturation of surface scintillators in the
shower core region begins to be significant.\\
\indent
The independent test of energy estimates can be done by the
detected zenith angle
distributions which have to be isotropic for different energy
thresholds. In Fig.~8 the corresponding detected distributions 
(symbols) are compared with statistically 
equivalent simulated isotropic distributions (lines).
The agreement of detected and simulated distributions at 
$E>3\cdot10^6$ GeV gives an 
additional support to the consistency of energy estimates in
 the whole measurement range. The anisotropic spectral behavior
at low energies ($E\sim(1-3)\cdot10^6$ GeV) is explained by the lack
of heavy nuclei at larger zenith angles in the detected
flux due to the applied shower selection criteria.\\
\begin{figure} 
\begin{center}
\includegraphics[width=7cm]{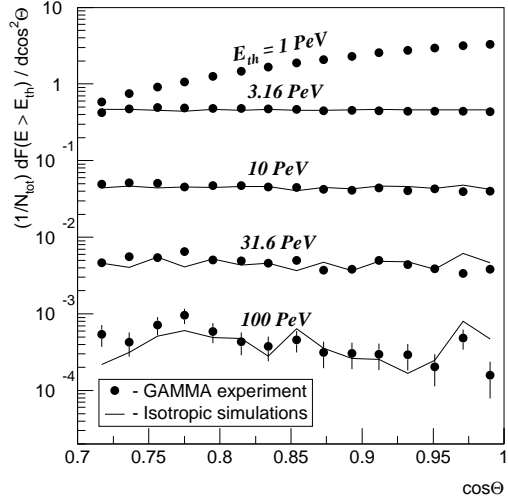}
\end{center}
\caption{Detected zenith angular distributions for different energy
thresholds (symbols). The lines are 
corresponding simulated isotropic distributions with the same
 statistics.}
\vspace{10mm}
\end{figure}
\indent
Using the aforementioned unbiased ($<5\%$) event-by-event method
of primary energy evaluation (10), we obtained the all-particle energy 
spectrum. Results
are presented in Fig.~9 (filled circle symbols, GAMMA07)
in comparison with the same spectra obtained
by the EAS inverse approach (line with shaded area, GAMMA06) from 
\cite{KAS07,G6a}
and our preliminary results (point-circle symbols, GAMMA05)
obtained using the 7-parametric event-by-event
method with the shower core selection criteria $R<25$m and
$\theta<30^0$ \cite{G5b}.\\ 
\begin{figure} 
\begin{center}
\includegraphics[width=8cm]{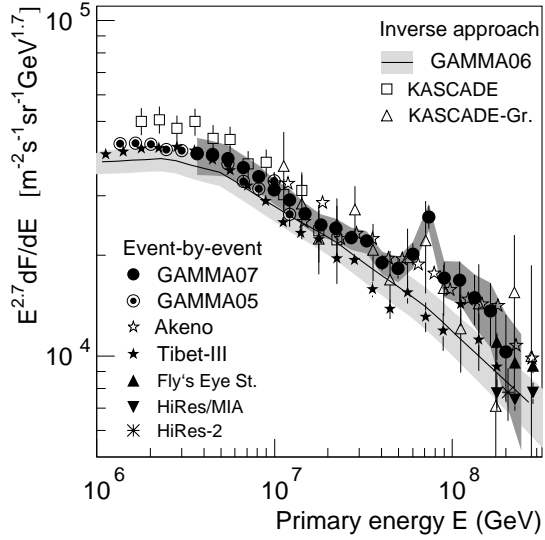}
\end{center}
\caption{All-particle energy spectrum in comparison with the results 
of EAS inverse
approach \cite{KAS07,G6a} and our preliminary data \cite{G5b}. 
The AKENO, Tibet-III, Fly's Eye Stereo,
Hires/MIA and Hires-2 data were taken from  
\cite{AKENO,Tibet,FlySt,HiResMIA,Hires2} respectively.}
\label{allps}
\vspace{10mm}
\end{figure}
\indent
It follows from our preliminary data \cite{G5b,G6b}, that 
the all-particle energy spectrum derived by event-by-event analysis 
with the 
multi-parametric energy
estimator (Section 3) depends only slightly on the interaction 
model (~QGSJET01 
\cite{QGSJET} or SIBYLL2.1 \cite{SIBYLL}~) and thereby, the errors of 
obtained spectra are mainly
 determined by the sum of statistical and systematic errors (7) presented 
in Fig.~9 by the dark shaded area.\\
\indent
Shower size detection threshold effects distort the all-particle spectrum 
in the range
of $E<(2-2.5)\cdot10^6$ GeV depending on the interaction model and determine the lower
 limit
$E_{\min}=3\cdot10^6$ GeV of the energy spectrum in Fig.~9 whereas the upper limit 
of the 
spectrum $E_{\max}\simeq(2-3)\cdot10^8$ GeV is determined by the saturation
 of our shower detectors which begins to be significant at 
$E_p>2\cdot10^8$ GeV and 
$E_{Fe}>4\cdot10^8$ GeV (see Fig.~4) for primary proton and $Fe$ nuclei.
The range of minimal systematic errors and biases is $(1-10)\cdot10^7$ GeV, where 
about $13\%$ and $10\%$ errors were attained (Fig.~3,4) for primary 
$H$ and $Fe$ nuclei respectively.\\
\indent
In Table~3 the numerical values of the obtained all-particle energy 
spectrum are 
presented along with statistical, total upper and lower 
errors according to (7) and corresponding number of detected events.
The energy spectra for low energy region (the first four lines) 
were taken from our data \cite{G5b} for the EAS selection criteria $R<25$m 
and $\theta<30^0$.\\
\begin{table}  
\caption{
All-particle energy spectrum ($d\Im/dE$) in units of
$(m^2\cdot sec\cdot sr\cdot GeV)^{-1}$
and corresponding statistical ($\Delta_{stat}$), total upper, 
($\Delta_+$) and total lower 
($\Delta_-$) errors and number of events ($N_{ev}$).
The first four lines represent our data \cite{G5b}
obtained for selection criteria $R<25$m and
$\theta<30^0$.}
\begin{center}
\begin{tabular}{|c|c|c|c|c|c|}
\hline
$E (PeV)$  & $d\Im/dE$  &$\Delta_{stat}$&$\Delta_+$&$\Delta_-$&   $N_{ev}$\\
\hline
\hline

$1.35 $ \cite{G5b}&$  0.1205E-11 $&$  0.11E-13 $& - & - &$   11550  $\\
$1.65 $ \cite{G5b}&$  0.7037E-12 $&$  0.77E-14 $& - & - &$    8374  $\\
$2.01 $ \cite{G5b}&$  0.4090E-12 $&$  0.54E-14 $& - & - &$    5769  $\\
$2.46 $ \cite{G5b}&$  0.2285E-12 $&$  0.36E-14 $& - & - &$    3924  $\\
$    3.00 $&$  0.1297E-12 $&$  0.52E-15 $&$  0.16E-13 $&$  0.20E-13$ & $59930$\\
$    3.67 $&$  0.7677E-13 $&$  0.37E-15 $&$  0.86E-14 $&$  0.108E-13$& $43157$ \\
$    4.48 $&$  0.4401E-13 $&$  0.25E-15 $&$  0.45E-14 $&$  0.57E-14$ & $30153$ \\
$    5.47 $&$  0.2524E-13 $&$  0.17E-15 $&$  0.24E-14 $&$  0.30E-14$ & $20993$ \\
$    6.69 $&$  0.1372E-13 $&$  0.12E-15 $&$  0.12E-14 $&$  0.15E-14$ & $13917$ \\
$    8.17 $&$  0.7506E-14 $&$  0.77E-16 $&$  0.62E-15 $&$  0.76E-15$ & $9290$ \\
$    9.97 $&$  0.3984E-14 $&$  0.51E-16 $&$  0.31E-15 $&$  0.37E-15$ & $5998$ \\
$   12.18 $&$  0.2166E-14 $&$  0.34E-16 $&$  0.17E-15 $&$  0.21E-15$ & $3986$ \\
$   14.88 $&$  0.1148E-14 $&$  0.23E-16 $&$  0.87E-16 $&$  0.104E-15$& $2573$ \\
$   18.17 $&$  0.619E-15  $&$  0.15E-16 $&$  0.45E-16 $&$  0.53E-16$ & $1692$ \\
$   22.20 $&$  0.350E-15  $&$  0.10E-16 $&$  0.25E-16 $&$  0.29E-16$ & $1170$ \\
$   27.11 $&$  0.1927E-15 $&$  0.69E-17 $&$  0.13E-16 $&$  0.15E-16$ & $781$ \\
$   33.12 $&$  0.1101E-15 $&$  0.47E-17 $&$  0.78E-17 $&$  0.88E-17$ & $542$ \\
$   40.45 $&$  0.556E-16  $&$  0.30E-17 $&$  0.42E-17 $&$  0.46E-17$ & $334$ \\
$   49.40 $&$  0.306E-16  $&$  0.20E-17 $&$  0.26E-17 $&$  0.27E-17$ & $227$ \\
$   60.34 $&$  0.199E-16  $&$  0.15E-17 $&$  0.18E-17 $&$  0.19E-17$ & $178$ \\
$   73.70 $&$  0.149E-16  $&$  0.12E-17 $&$  0.13E-17 $&$  0.14E-17$ & $164$ \\
$   90.02 $&$  0.572E-17  $&$  0.65E-18 $&$  0.70E-18 $&$  0.71E-18$ & $77$ \\
$   110.0 $&$  0.326E-17  $&$  0.44E-18 $&$  0.47E-18 $&$  0.47E-18$ & $54$ \\
$   134.3 $&$  0.184E-17  $&$  0.30E-18 $&$  0.31E-18 $&$  0.31E-18$ & $34$ \\
$   164.0 $&$  0.94E-18   $&$  0.19E-18 $&$  0.20E-18 $&$  0.20E-18$ & $22$ \\
$   200.3 $&$  0.40E-18   $&$  0.11E-18 $&$  0.12E-18 $&$  0.12E-18$ & $12$ \\
$   244.7 $&$  0.243E-18  $&$  0.81E-19 $&$  0.82E-19 $&$  0.82E-19$ & $7$ \\
\hline
\end{tabular}
\end{center}
\end{table}
\indent
The obtained energy spectrum agrees within errors with the KASCADE
\cite{KAS07},  
AKENO \cite{AKENO} and Tibet-III \cite{Tibet} data both in 
the slope and in the absolute 
intensity practically in the whole measurement range.
Looking at the 
experimental points we can unambiguously point out at the existence 
of an irregularity 
in the spectrum at the energy of $(6-8)\cdot10^7$ GeV. As it is seen from 
Figs~3 and 4, the energy estimator (10) has minimal biases ($\sim4-5\%$) 
and errors 
($\sim0.09-0.12$) at this energy.  With these errors the obtained bump has 
an apparently real nature.  If we fit all our other points in the
$(5-200)\cdot10^6$ GeV energy range by a smooth power-law spectrum, the bin
at $7.4\cdot10^7$ GeV exceeds this smooth spectrum by $4.0$ standard deviations.
The exact value for this significance of the bump depends somewhat on
the energy range chosen to adjust the reference straight line in Fig.~9,
but it lies in the range $(3.5-4.5)\sigma$.\\
\indent
We conservatively included the systematic errors in this estimate,
although they are not independent in the nearby 
points but correlated: the possible overestimation of the energy 
in one point cannot 
be followed by an underestimation in the neighboring point if their
 energies are 
relatively close to each other. Systematic errors can change slightly
 the general slope
of the spectrum but cannot imitate the fine structure and the existence 
of the bump.\\
\indent
The results from Fig.~7 show that in the range of "bump" energy 
($~7.4\cdot10^7$ GeV) the systematic errors can not increase 
significantly the flux. 
To test this hypothesis more precisely we tested the reconstruction
procedure for singular energy spectra with power indices $\gamma_p=1.5$
and $\gamma_p=4.5$ below and above the knee energy
 $~7.4\cdot10^7$ GeV in the  
$5\cdot10^7-10^8$ GeV energy range. Results are presented 
in Fig.~7 (star symbols) and show
that there are no significant discrepancies of 
reconstructed spectra observed, which stems from high accuracy
of energy reconstruction.\\
\indent
The detected shower sample in the bump energy region  
did not reveal any discrepancies with the showers from adjacent energy bins
within statistical errors
neither with respect to reconstructed shower core coordinates, zenith 
angular and $\chi^2$ distributions, nor with respect to 
$\xi=N_{ch}/N_{\mu}$ 
distribution. The only difference is that the average age of showers is 
increased from $\bar{s}=0.88\pm0.007$ at $E_0\simeq5\cdot10^7$ GeV up to 
$\bar{s}=0.93\pm0.01$ 
at $E_0\simeq10^8$ GeV, instead of the monotonic shower 
age decrease with energy 
increment, which is observed at a low energy region 
($E_0=3\cdot10^6-5\cdot10^7$ GeV).\\
\indent
It is necessary to note that some indications of the observed bump 
are also seen in KASCADE-Grande \cite{KAS07} (Fig.~9), TUNKA 
\cite{TUNKA} and Tibet-III \cite{Tibet} data but with 
larger statistical uncertainties at the level of 1.5-2 standard deviations.
Moreover, the locations of the bump in different 
experiments agree well with each other and with an expected knee energy for
$Fe$-like primary nuclei according to the rigidity-dependent
knee hypothesis \cite{G6,G6a}. However, the observed width ($\sim 20$\%
in energy) and height of the 
bump at the energy of $(6-8)\cdot10^7$ GeV, which exceeds 
by a factor of $\sim 1.5$
($\sim$ 4 standard deviations) the best fit 
straight line fitting all points above $5\cdot10^6$ GeV in Fig. 9, 
are difficult to 
describe
in the framework of the conventional model of cosmic ray origin \cite{Hill}.\\
\indent
As it will be shown below (Section 5, Fig.~10,11)
the detected EAS charged particle ($N_{ch}$) and muon size ($N_{\mu}$) spectra 
\cite{G6,G6a} 
independently indicate the existence of this bump just for the obtained 
energies
and as it follows from the behavior of shower age parameter versus
shower size \cite{G6,G6a}, the bump at energy $\sim7.4\cdot10^7$ GeV
 is likely formed
completely from $Fe$ nuclei.
\section{Possible origin of irregularities}
\indent
Irregularities of all-particle energy spectrum in the knee region 
are observed practically in all measurements  \cite{AKENO,KAS07,G6}
and are explained by both the rigidity-dependent knee hypothesis 
and contribution of pulsars in the Galactic cosmic ray flux
\cite{PB,Erlykin,Prot}. Two these approaches approximately
describe the all-particle spectrum in $(1-100)\cdot10^6$ GeV energy 
region.
However, the observed
bump in Fig.~9 at energies $\sim7.4\cdot10^7$ GeV both directly points out the 
presence
of additional component in the primary nuclei flux and displays 
a very flat  
($\gamma_{p}\sim0-2$) energy spectrum before a cut-off 
energy $E_c\simeq8\cdot10^7$ GeV.\\
\indent
It is known \cite{G6,G6a} that rigidity-dependent primary
energy spectra can not describe quantitatively the phenomenon of ageing
of EAS at energies $(5-10)\cdot10^7$ GeV
that was observed in the most mountain-altitude 
experiments \cite{G6,MAK,NORIKURA}. It is reasonable to assume that
an additional flux of heavy
nuclei ($Fe$-like) is responsible for the bump at these energies. Besides, 
the sharpness of the bump (Fig.~9) points out   
at the local origin of this flux from compact objects 
(pulsars) \cite{Erlykin,Prot}.\\
\indent
The test of this hypothesis we carried out using parameterized
inverse approach
\cite{G5a,G6,G6a} on the basis of GAMMA facility EAS database and 
hypothesis of two-component origin of cosmic ray flux: 
\begin{figure} 
\begin{center}
\includegraphics[width=7cm]{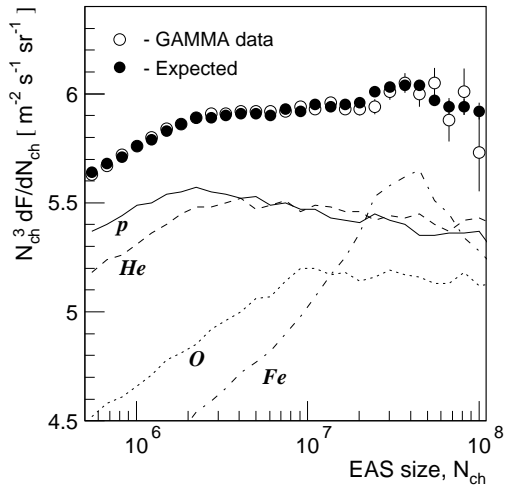}
\end{center}
\caption{
EAS size spectra detected by the GAMMA facility (empty symbols) 
and corresponding
expected spectra (filled symbols) computed in the framework of 
the SIBYLL2.1 interaction model and
2-component parametrization of primary spectra (11). 
The lines correspond to expected 
size spectra computed for each of primary nuclei.}
\vspace{10mm}
\end{figure}
\begin{figure} 
\begin{center}
\includegraphics[width=7cm]{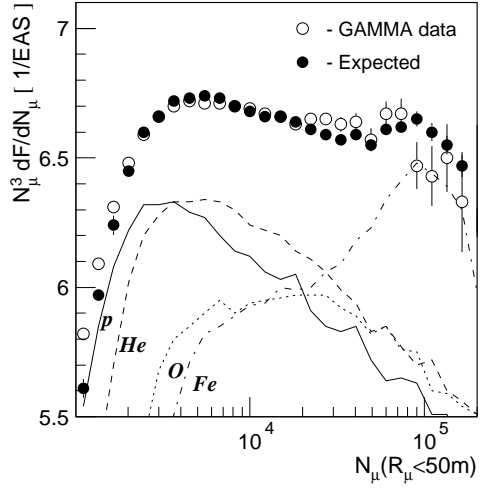}
\end{center}
\caption{The same as Fig.~10 for truncated EAS muon size spectra.}
\vspace{10mm}
\end{figure}
\begin{equation}
F_A(E)=\Phi_G(A)\Big( E_k^{-\gamma_1}
\Big(\frac{E}{E_k}\Big)^{-\gamma}+P_A(E)\Big)
\end{equation}
where $P_H=P_{He}=P_O=0$ and
\begin{displaymath}
P_{Fe}(E)=\Phi_P(Fe)\cdot E_{c,Fe}^{-\gamma_{1p}}
\Big(\frac{E}{E_{c,Fe}}\Big)^{-\gamma_p}.
\end{displaymath}
The first term in the right hand side of expression (11)
(so called Galactic component)
is the power law energy spectra
with rigidity-dependent knees at energies $E_k=E_R\cdot Z$ and power indices 
$\gamma=\gamma_1$ and 
$\gamma=\gamma_2$ for  $E\le{E_k}$ and $E>E_k$ respectively,  
and the second term 
(so called "pulsar component") is an
additional power law energy spectrum with cut-off energies $E_{c,Fe}$ and
power indices  $\gamma_p=\gamma_{1,p}$ and 
$\gamma_p=\gamma_{2p}$ for  $E\le{E_{c,Fe}}$ and $E>E_{c,Fe}$ 
respectively.\\
\indent
The scale factors 
$\Phi_G(A)$ and $\Phi_P(A)$ along with particle rigidity $E_R$, 
cut-off energy $E_c(A)$ and power indices 
$\gamma_1,\gamma_2,\gamma_{1p},\gamma_{2p}$ were estimated using 
combined approximation
method \cite{G5a,G6,G6a} for two examined shower spectra showed in 
Figs.~10,11 
(empty symbols):
EAS size spectra, $d F/d N_{ch}$ (Fig.~10) and EAS muon truncated size 
spectra, 
$d F/d N_{\mu}$ (Fig.~11)
detected by the GAMMA facility using shower core selection criteria 
$\theta<30^0$ and $r<50$m \cite{G6,G6a}. We did not consider
the $H$, $He$ and $O$ pulsar components
to avoid a large number of unknown parameters and corresponding 
mutually compensative pseudo solutions \cite{Pseudo} 
for the Galactic components.\\ 
\indent
The folded (expected) shower spectra (filled symbols in Figs 10,11) were 
computed on the basis 
of parametrization (11) and CORSIKA EAS simulated data set \cite{G6a,G6} 
for the $A\equiv H,He,O$ and $Fe$ primary nuclei to evaluate 
the kernel functions 
of corresponding integral equations \cite{G6,G6a}. 
The computation method, 
was completely the same as was performed in the combined 
approximation analysis \cite{G6,G6a}. The initial values of 
spectral parameters for
Galactic component were taken from \cite{G6a,G6} as well.
In Figs.~10,11 we also presented the derived expected elemental shower 
spectra (lines)
for primary $H,He,O$ and $Fe$ nuclei respectively.\\
\indent
The parameters of two-component primary energy spectra (11) derived from
the $\chi^2$ goodness-of-fit test 
of shower spectra $d F/d N_{ch}$ and $d F/d N_{\mu}$ 
are presented in Table~4.\\
\begin{table}  
\caption{\label{tab:table4}
Parameters of the primary energy spectra (11) derived
from combined approximations of detected shower spectra. 
The scale factors $\Phi_{G,P}(A)$ have units of 
(m$^2\cdot$ s $\cdot$ sr $\cdot$ TeV)$^{-1}$.
The particle rigidity $E_R$ and cut-off energies $E_c$ 
are shown in PV and PeV units respectively.}
\begin{center}
\begin{tabular}{|c|c|c|}
\hline
Param. & G-component&P-component\\
\hline
\hline
$\Phi(H ) $ & $.102\pm.003$&$-$\\
$\Phi(He)$& $.094\pm.022$&$-$\\
$\Phi(O) $& $.032\pm.015$&$-$\\
$\Phi(Fe)$& $.021\pm.006$&$(.29\pm.08)\cdot10^{-7}$\\
$\gamma_1$ & $2.68\pm.005$& $1.05\pm.5$\\
$\gamma_2$ & $3.29\pm.045$& $4.5\pm.4$\\
$E_R      $    & $2.59\pm.15$  &         $-$\\
$E_{c,Fe}$    &         $-$           & $76.9\pm1.5$\\
\hline
\end{tabular}
\end{center}
\vspace{5mm}
\end{table}
\indent
Resulting expected energy spectra $F_A(E)$ for 
the Galactic $H,He,O$ and $Fe$ nuclei (thin lines)
along with the all-particle spectrum $\sum_A{F_A(E)}$ 
(bold line with shaded area) are presented in Fig.~12.
The thick dash-dotted line corresponds to derived energy spectra
of the additional  $Fe$ component (the second term in the right hand side
of expression (11)). The all-particle
energy spectrum obtained on the basis of the GAMMA EAS data and event-by-event 
multi-parametric energy
evaluation method (Section~4, Fig.9) are shown in Fig.12 (symbols) as 
well.\\
\begin{figure} 
\begin{center}
\includegraphics[width=8cm]{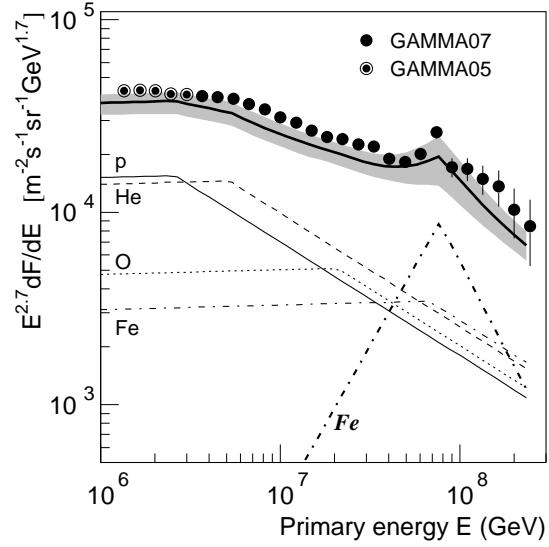}
\end{center}
\caption{All-particle primary energy spectrum (symbols)  
and expected energy spectra (lines and shaded area)   
derived from EAS inverse problem solution for $p,He,O$ and $Fe$ primary nuclei
using 2-component parametrization (11). The thin lines are 
the energy spectra of Galactic $H,He,O$ and $Fe$ components.
The thick dash-dotted line is an additional  $Fe$ component from
compact objects.}
\vspace{10mm}
\end{figure}
\indent
It is seen, that the shape of 2-component all-particle spectrum 
(bold line with shaded area)
calculated with parameters taken from the fit of EAS size spectra 
agrees within the 
errors with the results of event-by-event analysis (symbols)
that points out at the consistency of applied spectral parametrization 
(11) with GAMMA data.\\
\indent
Notice that the flux of derived additional $Fe$ component turned out to be 
about $0.5-0.6\%$ of the total $Fe$ flux for primary energies $E>10^6$ GeV.
This result agrees with expected flux of polar cap component 
\cite{PB}.\\
\begin{figure} 
\begin{center}
\includegraphics[width=7cm]{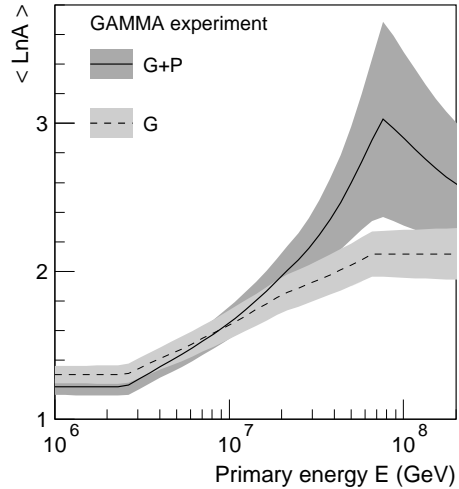}
\end{center}
\caption{
Average logarithm primary nuclei mass number derived from
rigidity-dependent primary energy spectra \cite{G6,G6a} 
(dashed line) and 2-component model prediction (11)
taking into account additional pulsar component (solid line).}
\vspace{10mm}
\end{figure}
\indent
The dependence of average nuclear mass number are presented 
in Fig.~13 
for two primary nuclei flux composition models: one-component model,
where the power law energy spectra 
of primary nuclei have rigidity-dependent knees at particle rigidity 
$E_R\sim2500$ GeV/Z \cite{G6,G6a}
(so called Galactic component, dashed line) and two-component model 
(solid line), where
additional pulsar (P) component was included according to 
parametrization (11) and data 
from Table~4 with very 
flat power index ($\gamma_{1p}\sim1$) before cut-off energy $E_{c,Fe}$. 
The shaded area in Fig.~13 show
the ranges of total (systematic and statistical) errors. 

\section{Conclusion}  
\indent
The multi-parametric event-by-event method (Sections 3,4) provides the 
high accuracy for 
the energy evaluation of primary cosmic ray nuclei $\sigma(E)\simeq10-15\%$
regardless of the nuclei mass (biases$<5\%$) in the $3-200$ PeV energy region. 
Using this method the all-particle energy spectrum in the knee region 
and above has 
been obtained (Fig.~9, Table~3) using the EAS database from the GAMMA 
facility. The results are obtained for the SIBYLL2.1 interaction model.\\
\indent
The all-particle energy spectrum in the range of statistical 
and systematic errors agrees with the same spectra obtained
using the EAS inverse approach \cite{G5a,KAS07,G6} in the $3-200$ PeV 
energy range.\\
\indent
The high accuracy of energy evaluations and small statistical errors point out at the existence of an irregularity (`bump') in the 
$60-80$ PeV primary energy region.\\
\indent
The bump can be described by 2-component model (parametrization (11)),
of primary cosmic ray origin,
where additional (pulsar) $Fe$ component are included with very flat 
power law energy spectrum 
($\gamma_{1p}\sim1\pm0.5$) before cut-off energy $E_{c,Fe}$ (Fig.~13, 
Table~4). 
At the same time, the EAS inverse problem solutions for energy spectra of 
pulsar component forces the solutions for the slopes of Galactic component
above the knee to be steeper (Table~4), that creates a problem of 
underestimation 
of all-particle energy spectrum in the range of HiRes \cite{HiResMIA} and 
Fly's Eye \cite{FlySt} data at $E>200$ PeV. 
From this viewpoint the underestimation ($10-15\%$) of the  
all-particle energy spectrum (bold solid line in Fig.~12) in the range 
of $E>200$ PeV
can be compensated by the expected extragalactic component \cite{Gaisser}.\\
\indent
Though we cannot reject the stochastic nature of the bump completely,
our examination of the systematic uncertainties of the applied method
lets us believe that they cannot be responsible for the observed feature.
The indications from other experiments mentioned in this paper provide
the argument for the further study of this interesting energy region.      
\section*{Acknowledgments} 
\indent
We are grateful to all our colleagues at the Moscow P.N.Lebedev 
Physical Institute and the Yerevan Physics Institute who took part 
in the development and exploitation of the GAMMA array. We thank 
Peter Biermann for fruitful discussions. \\
\indent
This work has been partly supported by research Grant No. 090 
from the Armenian government, by the RFBR grant 07-02-00491 in Russia, 
and by the "Hayastan" All-Armenian Fund and the ECO-NET project 12540UF.

\end{document}